\newif\ifdoubleblind
\newif\ifacm
	
\ifacm
	\documentclass[sigconf]{acmart}
\else
	\documentclass[conference]{IEEEtran}
\fi

\usepackage{amsmath}

\usepackage{flushend}
\usepackage{multirow}
\usepackage{tabularx,booktabs}
\usepackage{makecell}
\usepackage{xspace}

\usepackage[binary-units=true]{siunitx}
\usepackage{tikz}
\usetikzlibrary{positioning}

\ifacm
	\usepackage[nolist]{acronym}
\usepackage{booktabs} 
\usepackage{subfig}
\usepackage{balance}

\usepackage{array}
\usepackage{graphicx}
\usepackage{wasysym}
\usepackage{xparse}

\usepackage{makecell}
\newcolumntype{Y}{>{\centering\arraybackslash}X}

\hypersetup{draft}

\renewcommand\footnotetextcopyrightpermission[1]{} 
\else
	\usepackage{graphicx}
\usepackage{epstopdf}
\usepackage{standalone}
\usepackage[nolist ]{acronym}
\usepackage{tcolorbox}
\usepackage{pdfcomment}
\usepackage{hologo}

\usepackage{xstring}

\usepackage[utf8]{inputenc}
\usepackage{marvosym}

\usepackage{algorithm}
\usepackage{algpseudocode}

\usepackage{psfrag}
\usepackage{graphicx}

\usepackage{rotating}

\renewcommand{\baselinestretch}{1}

\usepackage[caption=false,font=footnotesize]{subfig}
\usepackage{stfloats}

\fi

\begin{document}

\newcommand{\paperTitle}{Flying Robots for Safe and Efficient Parcel Delivery Within the COVID-19 Pandemic}
\newcommand{\paperAuthors}{Manuel Patchou, Benjamin Sliwa, and Christian Wietfeld}
\newcommand{\paperEmails}{$\{$Manuel.Mbankeu, Benjamin.Sliwa, Christian.Wietfeld$\}$@tu-dortmund.de}

\newcommand\single{1\textwidth}
\newcommand\double{.48\textwidth}
\newcommand\triple{.32\textwidth}
\newcommand\quarter{.24\textwidth}
\newcommand\singleC{1\columnwidth}
\newcommand\doubleC{.475\columnwidth}

\newcommand{\figurePadding}{0pt}
\newcommand{\figureTopPadding}{\figurePadding}
\newcommand{\figureBottomPadding}{\figurePadding}
\newcommand\red[1]{\colorbox{red}{#1}}

\newcommand\tikzFig[2]
{
	\begin{tikzpicture}
		\node[draw,minimum height=#2,minimum width=\columnwidth,text width=\columnwidth,pos=0.5]{\LARGE #1};
	\end{tikzpicture}
}

\newcommand{\dummy}[3]
{
	\begin{figure}[b!]  
		\begin{tikzpicture}
		\node[draw,minimum height=6cm,minimum width=\columnwidth,text width=\columnwidth,pos=0.5]{\LARGE #1};
		\end{tikzpicture}
		\caption{#2}
		\label{#3}
	\end{figure}
}

\newcommand\pos{h!tb}

\newcommand{\basicFig}[7]
{
	\begin{figure}[#1]  	
		\vspace{#6}
		\centering		  
		\includegraphics[width=#7\columnwidth]{#2}
		\caption{#3}
		\label{#4}
		\vspace{#5}	
	\end{figure}
}
\newcommand{\fig}[4]{\basicFig{#1}{#2}{#3}{#4}{0cm}{0cm}{1}}
\newcommand{\figw}[5]{\basicFig{#1}{#2}{#3}{#4}{0cm}{0cm}{#5}}

\newcommand\sFig[2]{\begin{subfigure}{#2}\includegraphics[width=\textwidth]{#1}\caption{}\end{subfigure}}
\newcommand\vs{\vspace{-0.3cm}}
\newcommand\vsF{\vspace{-0.4cm}}

\newcommand{\subfig}[3]
{%
	\subfloat[#3]%
	{%
		\includegraphics[width=#2\textwidth]{#1}%
	}%
	\hfill%
}

\newcommand\circled[1] 
{
	\tikz[baseline=(char.base)]
	{
		\node[shape=circle,draw,inner sep=1pt] (char) {#1};
	}\xspace
}

\begin{acronym}
	\acro{LIMoSim}{LIghtweight ICT-centric Mobility Simulation}
	\acro{LIMITS}{LIghtweight Machine learning for IoT System}
	\acro{ITS}{Intelligent Transportation System}
	\acro{UAV}{Unmanned Aerial Vehicle}
	\acro{OSM}{OpenStreetMap}
	\acro{ns-3}{Network Simulator 3}
	\acro{HIL}{Hardware-in-the-Loop}
	\acro{DDNS}{Data-Driven Network Simulation}
	\acro{RAT}{Radio Access Technology}
	\acro{REM}{Radio Environmental Map}
	\acro{TSP}{Travelling Salesman Problem}
	\acro{3GPP}{3rd Generation Partnership Project}
	\acro{5GAA}{5G Automotive Association}
	\acro{QoS}{Quality of Service}
	\acro{IoT}{Internet of Things}
	\acro{MNO}{Mobility Network Operator}	
	\acro{FSTSP}{Flying Sidekick Travelling Salesman Problem}
	\acro{MIP}{Mixed Integer Programming}
	\acro{A2A}{Air-to-Air}
	\acro{A2G}{Air-to-Ground}
	\acro{CUSCUS}{CommUnicationS-Control distribUted Simulator}
	\acro{FL-AIR}{Framework libre AIR}
	\acro{IPC}{Inter-Process Communications}
	\acro{LTE}{Long Term Evolution}
	\acro{PDR}{Packet Delivery Ratio}
	\acro{MPTCP}{Multi-Path Transmission Control Protocol}
	\acro{IPD}{Inter-Point Distance}
	\acro{C-V2X}{Cellular Vehicle-to-Everything}
	\acro{WAVE}{Wireless Access for Vehicular Environment}
	\acro{LOS}{Line of Sight}
	\acro{eNB}{evolved NodeB}
	\acro{SPS}{Semi-Persistent Scheduling}
	\acro{CAM}{Cooperation Awareness Message}
\end{acronym}

\title{\paperTitle}

\ifacm
\else

	\title{\paperTitle}

	\ifdoubleblind
	\author{\IEEEauthorblockN{\textbf{Anonymous Authors}}
		\IEEEauthorblockA{Anonymous Institutions\\
			e-mail: Anonymous Emails}}
	\else
	\author{\IEEEauthorblockN{\textbf{\paperAuthors}}
		\IEEEauthorblockA{Communication Networks Institute,	TU Dortmund University, 44227 Dortmund, Germany\\
			e-mail: \paperEmails}}
	\fi
	
	\maketitle

\fi




%
%
\def\COPYRIGHTYEAR{2021}
\def\CONFERENCE{15th IEEE International Systems Conference (SYSCON) 2021} 

\def\bibtex
{
@InProceedings\{Patchou/etal/2021a,
	author    = \{Manuel Patchou and Benjamin Sliwa and Christian Wietfeld\},
	title     = \{Flying Robots for Safe and Efficient Parcel Delivery Within the COVID-19 Pandemic\},
	booktitle = \{IEEE International Systems Conference (SYSCON)\},
	year      = \{2021\},
	address   = \{Vancouver, Canada\},
	month     = \{Apr\},
\}
}
\ifx\CONFERENCE\VOID
\def\conferencenotice{Submitted for publication}
\def\copyrightnotice{}
\else
\ifx\DOI\VOID
\def\conferencenotice{Accepted for presentation in: \CONFERENCE}	
\else
\def\conferencenotice{Published in: \CONFERENCE\\DOI: \href{http://dx.doi.org/\DOI}{\DOI}

}
\fi
\def\copyrightnotice{
	\copyright~\COPYRIGHTYEAR~IEEE. Personal use of this material is permitted. Permission from IEEE must be obtained for all other uses, including reprinting/republishing this material for advertising or promotional purposes, collecting new collected works for resale or redistribution to servers or lists, or reuse of any copyrighted component of this work in other works.
}
\fi

\def\overlayimage{%
	\begin{tikzpicture}[remember picture, overlay]
	\node[below=5mm of current page.north, text width=20cm,font=\sffamily\footnotesize,align=center] {\conferencenotice \vspace{0.3cm} \\ \pdfcomment[color=yellow,icon=Note]{\bibtex}};
	\node[above=5mm of current page.south, text width=15cm,font=\sffamily\footnotesize] {\copyrightnotice};
	\end{tikzpicture}%
}
\overlayimage

\begin{abstract}
	
%
%
The integration of small-scale \acp{UAV} into \acp{ITS}  will empower novel smart-city applications and services.
%
%
After the unforeseen outbreak of the COVID-19 pandemic, the public demand for delivery services has multiplied. Mobile robotic systems inherently offer the potential for minimizing the amount of direct human-to-human interactions with the parcel delivery process.
%
%
The proposed system-of-systems consists of various complex aspects such as assigning and distributing delivery jobs, establishing and maintaining reliable communication links between the vehicles, as well as path planning and mobility control. 
In this paper, we apply a system-level perspective for identifying key challenges and promising solution approaches for modeling, analysis, and optimization of \ac{UAV}-aided parcel delivery. We present a system-of-systems model for \ac{UAV}-assisted parcel delivery to cope with higher capacity requirements induced by the COVID-19.
%
%
To demonstrate the benefits of hybrid vehicular delivery, we present a case study focusing on the prioritization of time-critical deliveries such as medical goods. The results further confirm that the capacity of traditional delivery fleets can be upgraded with drone usage. Furthermore, we observe that the delay incurred by prioritizing time-critical deliveries can be compensated with drone deployment. Finally, centralized and decentralized communication approaches for data transmission inside hybrid delivery fleets are compared.

\end{abstract}

\maketitle

\section{Introduction}

%
%
The integration of small-scale \acp{UAV} into future \ac{ITS} \cite{Menouar/etal/2017a} has been an emerging topic in recent years. Due to the unique mobility potential of these airborne vehicles, significant efficiency improvements in applications such as aerial traffic monitoring \cite{Sliwa/etal/2019b} and parcel delivery \cite{Patchou/etal/2019a} have been anticipated.
%
%
%
Despite promising initial case studies, the existing approaches have been dominated by academic analyses.
%
%
However, the public interest in implementing mobile robotics-based delivery systems has been massively increased after the outbreak of the COVID-19 pandemic \cite{Chamola/etal/2020a}.

\fig{}{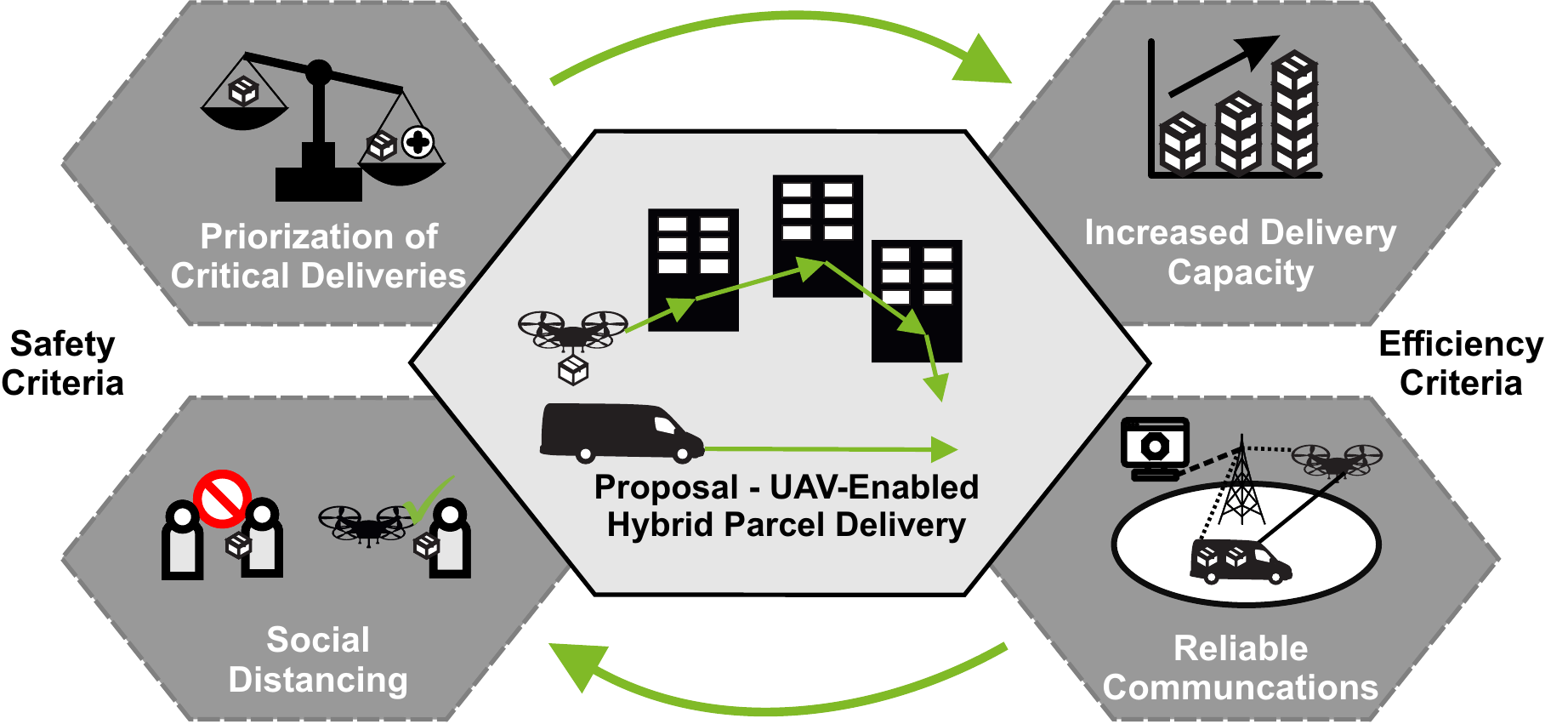}{Overview of the upgraded challenges faced by the parcel delivery sector during the COVID-19 pandemic.}{fig:scenario}
%
%
In this paper, we present a system-of-systems approach for implementing the various aspects of hybrid vehicular delivery systems, which consist of collaborating vehicles on the ground and in the air. Our work is motivated by the following core observation in the context of COVID-19, which are also illustrated in Fig.\ref{fig:scenario}:
\begin{itemize}
	%
	%
	\item \textbf{Human-to-human interactions} should be minimized in order to avoid the further spread of the virus. Mobile robotic vehicles can serve as non-infective delivery agents.
	%
	%
	\item \textbf{Capacity enhancement} of delivery services is required to cope with the increasing demand for delivered goods.
	%
	%
	\item \textbf{Time-critical goods} such as medical equipment need to be prioritized over standard consumer goods. Aerial vehicles are not affected by traffic jams and can be utilized for rapid shipment even in highly crowded inner-city scenarios.
	\item \textbf{Reliable communications} are required for the safe operation and monitoring of the hybrid delivery fleet and networking of the delivery truck and the \ac{UAV} while proposing command and control features.
\end{itemize}
Due to the resulting complexity, the systems thinking is applied to propose a solution.
%
%
The contributions are summarized as follows:
%
%
\begin{itemize}
	\item A \textbf{system-of-systems concept} for an optimized \ac{UAV}-enabled last-mile delivery platform.
	\item A \textbf{proof-of-concept study} in a real-world scenario to analyze capacity gains and suitable network approaches.
	\item The \textbf{prioritization of time-critical deliveries} such as medical goods and its effects on the delivery performance.
\end{itemize}

%
%
The remainder of the paper is structured as follows. After discussing the related work in Sec.~\ref{sec:related_work}, the proposed system model is presented in Sec.~\ref{sec:approach}. Afterwards, an overview of the methodological aspects is given in Sec.~\ref{sec:methods}. Finally, detailed results about conducted case studies are provided in Sec.~\ref{sec:results}.

\section{Related Work} \label{sec:related_work}

%
%
\textbf{\ac{UAV}-assisted delivery systems}:
Last-mile delivery is often completed by trucks driving over relatively short distances in an urban setting. The number and the variety of shipments, the complexity, and frequent congestion of the urban environment pose a challenge in terms of reliability and efficiency.
The inclusion of \acp{UAV} in the last mile delivery fleet is now under active consideration and testing.

A novel approach in this direction lets the delivery truck carry the drones when leaving the depot center. These are then dispatched along the route to autonomously deliver parcels up to a certain distance from the truck.
This approach requires finding an optimal route for the delivery truck, taking into account the mobility range of its associated delivery drones, which is limited by their battery capacity thereby leading to a new class of \emph{hybrid} \ac{TSP}.
%
%
This optimization problem was labeled the \ac{FSTSP} in one of its earliest formulations as a \ac{MIP} problem and presented with a heuristic solution approach in \cite{Murray2015}. The initial \ac{TSP} tour is iteratively updated by checking whether the next node shall be served by a drone and assigning it accordingly.
%
%
The work in \cite{Wang2017} derived several worst-case results on the hybrid routing problem with drones, such as the maximal savings, which can be obtained by their usage.
%
%
A dynamic pickup and delivery approach, allowing aerial vehicles to pick up packets
from a moving truck and autonomously complete deliveries, was investigated in \cite{Marinelli2018}. Comparing this approach to the more traditional one, where the drones deliveries may only be started when the truck stops to complete a delivery, shows further performance improvements.

%
%
Drone aided parcel delivery must be, like all autonomous vehicles applications, guaranteed sufficient networking resources for monitoring, safety, and operational purposes. As such, providing cellular connectivity to low altitude \acp{UAV} has gathered increasing interest from the industry \cite{3GPP/2017a} \ac{3GPP}.
%
%
The variation in resource availability is also of importance, for example, forecasting the data rate along vehicular trajectories. Use cases and standard mechanisms proposals for \emph{predictive \ac{QoS}} are presented in \cite{5GAA/2020a}.
%
%
An overview of distinct communication \ac{QoS} requirements for \ac{UAV} applications, differentiated, among other criteria, by the nature of transmitted data  is given in \cite{Zeng/etal/2019a}.
%
%
An additional \ac{UAV} enabled application is presented in \cite{Zhou/etal/2015a}, where a ground vehicular network is enhanced with multiple \acp{UAV} forming an aerial sub-network through \ac{A2A} and \ac{A2G} links.

%
%
\textbf{Simulation of \ac{UAV} networks}:
%
%
Simulation is still the leading approach for validating and evaluating vehicular network solutions, as shown in \cite{Cavalcanti/etal/2018a}. Different simulation frameworks for \ac{UAV} networks have already been proposed.
%
%
\emph{\ac{CUSCUS}}\cite{Zema/etal/2018a} provides interconnection of ns-3 and \ac{FL-AIR} based on Linux containers with some limitations regarding simulated network connectivity, as \ac{LTE}, is currently not supported.
%
%
\emph{FlyNetSim}\cite{Baidya/etal/2018a} is a \ac{HIL} focused platform, which follows a middleware-based approach to couple ns-3 with Ardupilot.
%
%
\ac{LIMoSim} \cite{Sliwa/etal/2019c} applies a shared codebase approach, thus avoiding \ac{IPC} overhead, to couple ns-3 with the \ac{LIMoSim} mobility kernel supporting hybrid combinations of ground and aerial vehicles.
A dedicated extension was provided in \cite{Patchou/etal/2019a} to bring native simulation support of \ac{UAV}-aided parcel delivery mobility and communications.

%
%
\textbf{Machine learning}:
%
%
\ac{DDNS} \cite{Sliwa/Wietfeld/2019c} is a novel concept for analyzing the end-to-end performance of anticipatory mobile networking systems while leveraging an accurate and fast network simulation, based on the observation of \emph{concrete environments} in combination with machine learning models. The generated network behavior's accuracy is comparable to what system-level simulations provide without the usual overhead associated to protocol simulation.
%
%
\cite{Thrane/etal/2020a} proposes a lightweight alternative to computationally expensive ray tracing evaluations often based on extensive knowledge of scenario topology. The model aided deep learning extracts radio propagation characteristics from aerial images of the receiver environment.
%
%
A comprehensive summary of machine learning methods and their different communication networks related applications is provided in\cite{Wang/etal/2020a}.

\section{Proposed System-of-Systems Model for UAV-Aided Parcel Delivery} \label{sec:approach}

In this section, we present a hybrid vehicular delivery service concept to cope with  increasing demand, which can be induced by special situations such as the COVID-19 pandemic, and we identify requirements and promising solution approaches.
%
%
\begin{figure*}[]
	\vspace{0cm}
	\centering
	\includegraphics[width=1.0\textwidth]{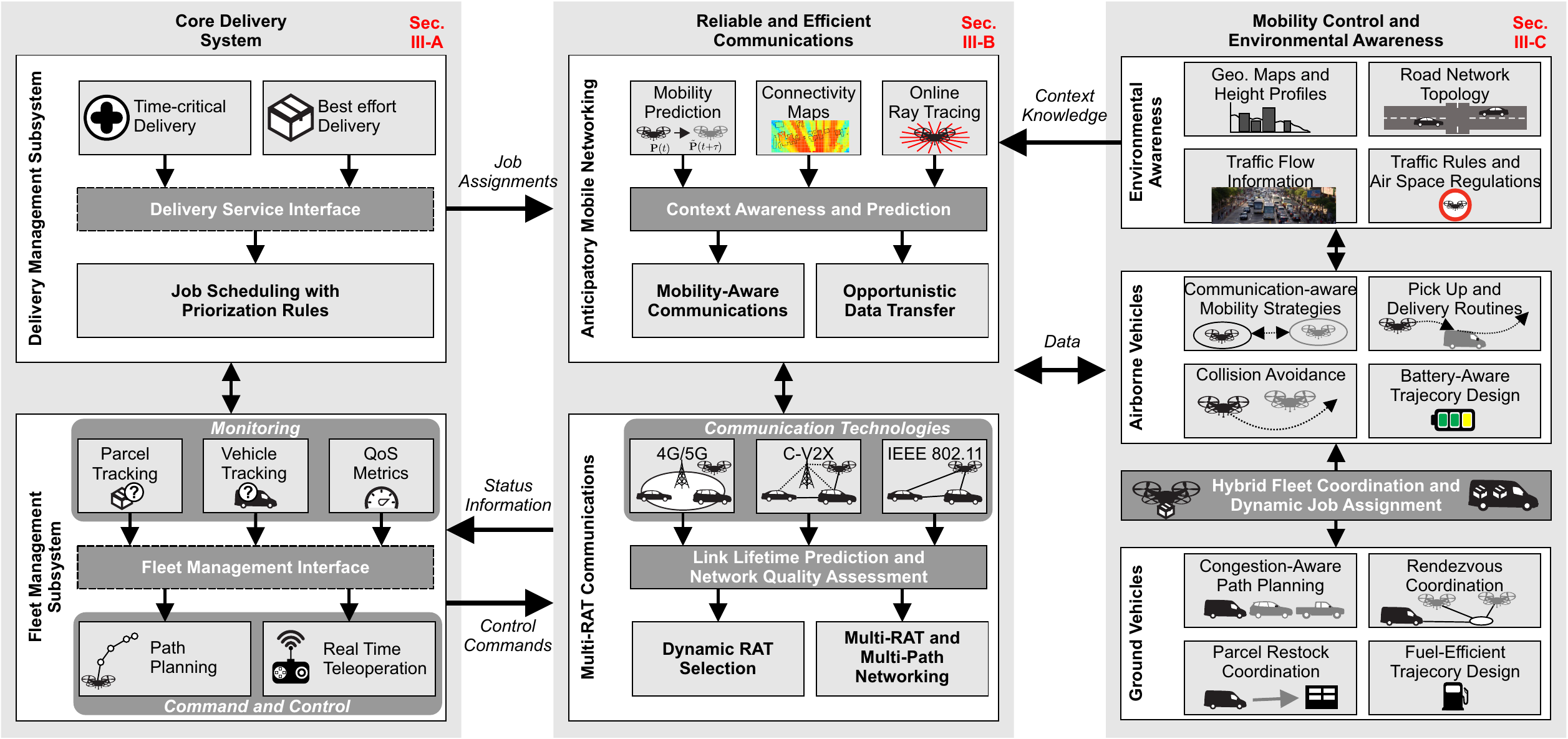}
	\caption{System-of-systems architecture model of hybrid vehicular delivery service implementation. The delivery fleet comprises cooperating vehicles on the ground and in the air.}
	\label{fig:architecture}
	\vspace{0cm}
\end{figure*}
The overall system-of-systems architecture model of the proposed hybrid vehicular delivery service is shown in Fig.~\ref{fig:architecture}. It consists of three closely related building blocks which are explained in further details in the following paragraphs.
%
%
%

%
%
\subsection{Core Delivery System} \label{sec:delivery}

%
%
The core delivery system provides the interfaces of service which can be used to provision and interact with consumer facing applications or internal service provider tools. It is further divided into two specialized subsystems for delivery and fleet management.

%
%
\textbf{Delivery management subsystem}:
%
%
One major requirement of the service is a channel through which delivery jobs can be created. The service must offer interfaces allowing external provisioning of deliveries with additional metadata specification to allow categorization.
These interfaces must support common communication protocols and standards for flexible integration.
%
%
The aforementioned job categorization is used to classify delivery requests into \emph{time-critical} (e.g., for medical applications) and \emph{best-effort} deliveries.
%
%
This classification is used by the job \emph{scheduling} algorithm to increase service performance and meet the deadlines of time-critical deliveries by respecting a set of prioritization rules for the final job assignment process.
%
%
%

%
%
\textbf{Fleet management subsystem}:
%
%
Operating such a decentralized system requires \emph{monitoring} in order to
assess the system's performance, detect optimization potential or bottlenecks, and provide tracking features. One very common feature directed to consumers in last-mile delivery is parcel tracking, allowing them to follow the journey of their parcel. Tracking of delivery agents is important to the service provider and is provided as well.

%
%
In addition to the monitoring feature, remote \emph{command and control} is needed to
constantly watch over the delivery fleet and promptly react to unforeseen events.
Control can be exerted in a soft way, using path planning to reroute delivery agents, by defining waypoints to which they autonomously navigate. A strict control mode is provided as well to allow taking full remote control of the delivery drones in urgent and delicate circumstances.

%
%
While path planning may only consist of sending waypoints amounting to a few bytes to the delivery \acp{UAV}, full vehicle teleoperation requires guaranteeable network performance for the end-to-end latency, data rate, and \ac{PDR}. As a consequence, the usage of drones as delivery agents has specific requirements which are discussed next.

\subsection{Reliable and Efficient Communications} \label{sec:communications}

The availability of a reliable and efficient means of communication between the different agents and the command and control service is one of the prerequisites for implementing \ac{UAV} enabled delivery services and autonomous vehicular applications in general. Taken from \cite{Zeng/etal/2019a}, general network requirements for typical \ac{UAV} applications are shown in the table Tab.~\ref{tab:uav-general}, while more network requirements for more specific \ac{UAV} use cases can be seen from
Tab.~\ref{tab:uav-pdp}.
%
%

\begin{table}[ht]
	\centering
	\caption{General requirements of typical \ac{UAV} applications}
	\begin{tabular}{llll}
		\toprule
		\textbf{Link  Data Type} & \textbf{Data Rate} & \textbf{Reliability} & \textbf{Latency}\\
		\midrule
		DL C\&C & 60-100~kbps & \(10^-3\)~PER & 50~ms \\
		UL C\&C & 60-100~kbps & \(10^-3\)~PER & -\\
		UL Application Data &  up to 50~Mbps & - & \makecell{Similar to \\terrestrial user} \\
		\bottomrule
	\end{tabular}

	\vspace{0.1cm}
	C\&C: Command and Control
	DL: Downlink,
	UL: Uplink,
	PER: Packet Error Rate
	\label{tab:uav-general}
\end{table}

%
%

\begin{table}[ht]
	\centering
	\caption{Network requirements for \ac{UAV} applications}
	\begin{tabular}{lll}
		\toprule
		\textbf{Application} & \textbf{Latency} & \textbf{Data rate (DL/UL)}\\		
		\midrule		
		Drone delivery & 500~ms & 300~Kbps / 200~Kbps \\
		Drone filming & 500~ms & 300~Kbps / 200~Kbps \\
		Access point & 500~ms & 300~Kbps / 200~Kbps \\
		Infrastructure inspection & 3000~ms & 300~Kbps / 10~Mbps \\
		Search and rescue & 500~ms & 300~Kbps / 6~Mbps \\		
		\bottomrule
		
	\end{tabular}

	\vspace{0.1cm}
	DL: Downlink,
	UL: Uplink
	\label{tab:uav-pdp}
\end{table}


%
%
%
For establishing and maintaining communication links with the delivery fleet, we propose the exploitation of two network optimization principles, which are further introduced in the following paragraphs.

%
%
\textbf{Anticipatory mobile networking}:
%
%
Opportunistic data transfer for delay-tolerant data allows saving network resources by transmitting large data chunks in a context-aware manner, preferably during \emph{connectivity hotspots} \cite{Sliwa/etal/2019d}. This improves power consumption, which is critical for battery-constrained devices. The context-aware transmission predicts the network quality based on the variation of network indicators and the vehicle's position over time. An advanced version uses machine learning models to produce more accurate predictions. Inversely, local routing decisions \cite{Sliwa/etal/2021a} can be made using these predictions, thereby preferring the routes with better network conditions.
%
%
Using these models on the delivery agents requires
rapid prototyping of the targeted systems and a suitable methodology to migrate high-level machine learning models to resource-constrained \ac{IoT} platforms.
\ac{LIMITS} \cite{Sliwa/etal/2020c} allows automating high-level machine learning tasks and automatically deriving \texttt{C/C++} implementations of trained prediction models that can be executed on real-world platforms as well as in system-level network simulations such as \ac{ns-3}.

%
%
\textbf{Multi-\ac{RAT}} allows utilizing redundant communication interfaces for network communications, thus freeing systems from a critical dependency on the network infrastructure of a sole provider.
This approach can be implemented within a defined communication technology standard, for example, by bundling the networks of multiple \acp{MNO} together to offer reliable network access, while even including complimentary base station deployments \cite{gueldenring2020}.
Technology switching, e.g. through joint usage of ad-hoc networks and cellular, harnessing the advantages of both centralized and decentralized network approaches, is also possible.
%
%
SKATES \cite{Gueldenring/etal/2020a} is a multi-link capable communication module that fully aggregates the available \acp{RAT} and offers such switching capabilities by using a \ac{MPTCP} Linux kernel to seamlessly distribute traffic over all available network interfaces without connection interruption. Connectivity is given so long as one interface still has network access. Such a device is a suitable candidate to equip delivery drones, and more generally, autonomously operating or remote-controlled agents.

\subsection{Mobility Control and Environmental Awareness} \label{sec:mobility}

%
%
Strong interdependency between environment, mobility, and communication calls for joint consideration of these aspects and their influence on one another.

%
%
\textbf{Environment}:
Knowledge about the static and mobile obstacles is of crucial importance for the collision avoidance and mobility control of autonomous \acp{UAV}. These obstacles can also act as attenuators which impact the radio propagation effects.
Importing map data from \ac{OSM} in a simulator offers real environment models with urban obstacles.
A further obstacle, often not considered in most network simulations, is the ground. We can account for \emph{terrain profile} in the simulation by extracting it from the \ac{OSM} database as well.
%
%
This accurate representation of real-world scenarios can be used to create \acp{REM} for a preparatory analysis for strategic service deployment.

%
%
\textbf{Mobility}:
The physical and logistical characteristics of the agents are a prominent factor in delivery performance and define their suitable usage in the fleet. The traditional
\emph{truck} has a high parcel capacity and is used to deliver multiple parcels in one tour. Its performance is, however, constrained by the road network rules and eventual traffic congestion.
A \emph{\ac{UAV}} is, on the contrary, free of such constraints as it can maneuver in three dimensions. Its capacity is, however, limited as it can only carry one parcel. The reach is limited as well, due to its battery capacity.
%
%
Combining these two vehicle types in a hybrid delivery fleet brings performance benefits when applying a joint path planning of aerial- and ground-based vehicles \cite{Patchou/etal/2019a}. An optimal path can be found by solving the resulting hybrid \ac{TSP} with drones.

A cross-layer approach that leverages knowledge from the mobility control routines for proactive optimization of the communication performance is envisaged.
On the one hand, \emph{mobility-aware communication} methods allow considering the estimations about the future mobility of the vehicles to proactively schedule data transmissions.
On the other hand, \emph{communication-aware mobility} approaches actively perform navigation while privileging routes with better network conditions estimates.
%
%
Reinforcement learning can be used to train an agent to this effect and create a complex navigation behavior that can combine or alternate priorities between taking the shortest path to the target and experiencing optimized network conditions along the way.

\section{Simulation-Based Performance Analysis} \label{sec:methods}

In this section, we present the methodological aspects of the simulation-based performance analysis.
%
%
In order to jointly analyze the mobility behavior of the different vehicles and the corresponding communication between the logical entities, we utilize the \ac{LIMoSim} \cite{Sliwa/etal/2019c} mobility simulator in combination with \ac{ns-3}.
%
%
%

%
%
\subsection{Parcel Delivery Simulation}
We use the \ac{LIMoSim} extension presented in \cite{Patchou/etal/2019a} to simulate last-mile  parcel delivery. The  en-route scheme was applied to allow drone dispatching and recovery along the truck's route without stopping due to its confirmed time gains. This framework transforms the delivery problem in a \ac{TSP}, which is then solved using an adequate algorithm.
The simulations are run while using different communication technologies. Two global approaches are compared for indicated usage in drone-aided parcel delivery communications.

\subsection{Prioritization Scheduling}
%
%
Deliveries may be prioritized differently, based on various criteria and factors. Even more so, during pandemics where medical supply is of great importance. We consider this fact by applying a simple scheme for prioritizing important deliveries, further labeled as medical over normal deliveries. The delivery list is divided between medical deliveries and normal ones. the medical delivery set is considered as a standalone problem and solved first. The last delivery of the medical delivery set is then used as the starting point for finding the optimal delivery path for the normal delivery set as shown in Fig~\ref{fig:prio}.
\figw{}{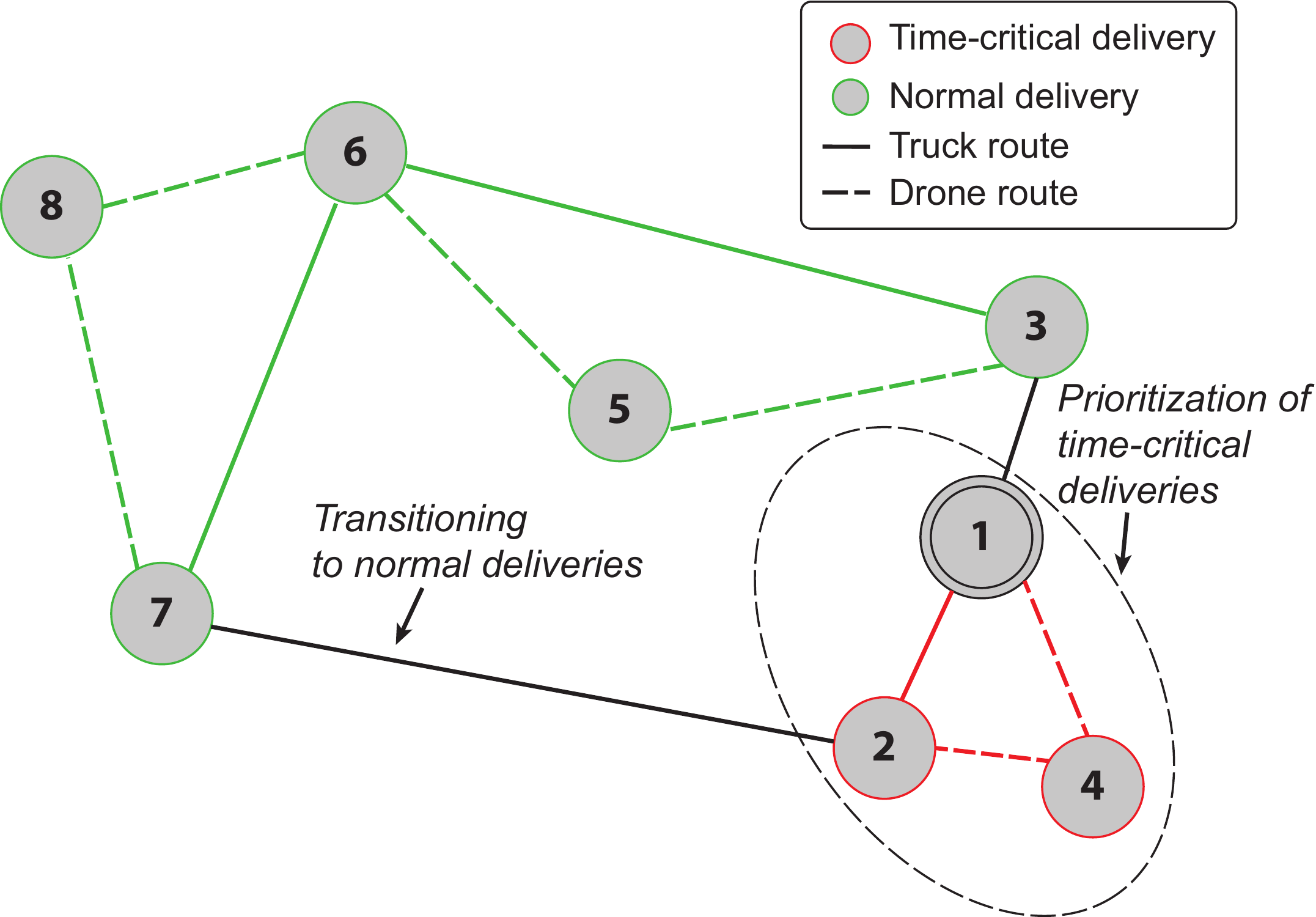}{Prioritization scheduling used for medical deliveries.}{fig:prio}{0.8}

%
%
To observe the performance of the delivery process, we consider the waiting time for each delivery. The communication technologies are evaluated using  performance metrics such as latency for responsiveness and \ac{PDR} for reliability.
%
%
The evaluations are performed within a suburban environment around a university campus. A map of the considered simulation scenario within \ac{LIMoSim} is shown in
Fig.~\ref{fig:simulation-world}.\\
Tab.~\ref{tab:parameters} summarizes the key simulation parameters.
\fig{}{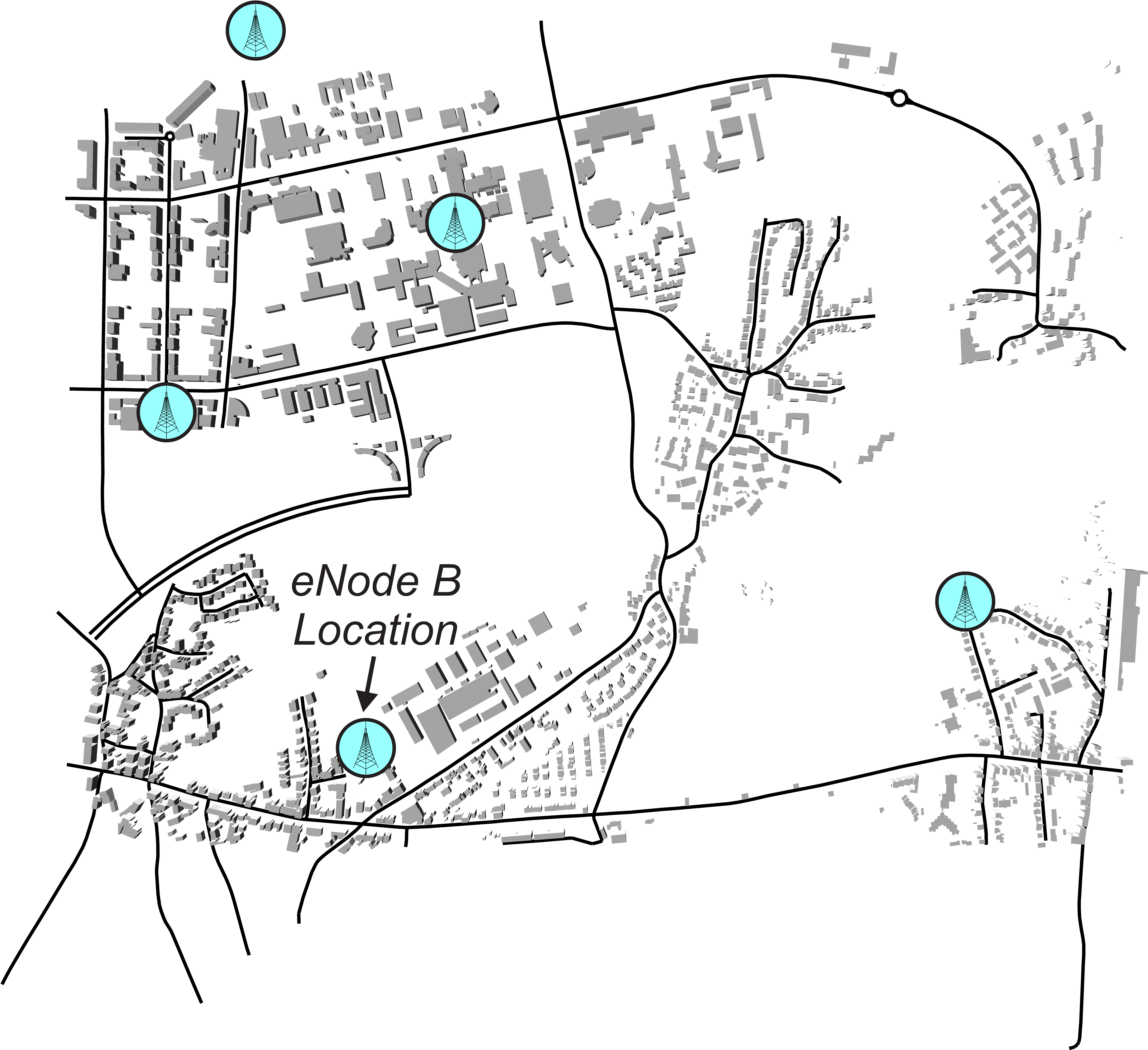}{Map of the evaluation scenario: The Dortmund university campus. (Map data: $\copyright$ OpenStreetMap Contributors, CC BY-SA).}{fig:simulation-world}
%
%
\newcommand{\entry}[2]{#1 & #2 \\}
\begin{table}[ht]
	\centering
	\caption{Default parameters of the evaluation setup}
	\begin{tabular}{ll}
		\toprule
		\textbf{Parameter} & \textbf{Value} \\

		\midrule
		
		\entry{Delivery sets}{50}
		\entry{Deliveries per set}{15}
		\entry{Medical Deliveries per Set}{5}
		\entry{UAV count}{\{~0,~1,~2,~3,~4,~5~\}}		
		\bottomrule
		
	\end{tabular}
	\label{tab:parameters}
\end{table}

\section{Results of the Proof-of-Concept Evaluation} \label{sec:results}

\begin{figure}[]
	\centering
	\subfig{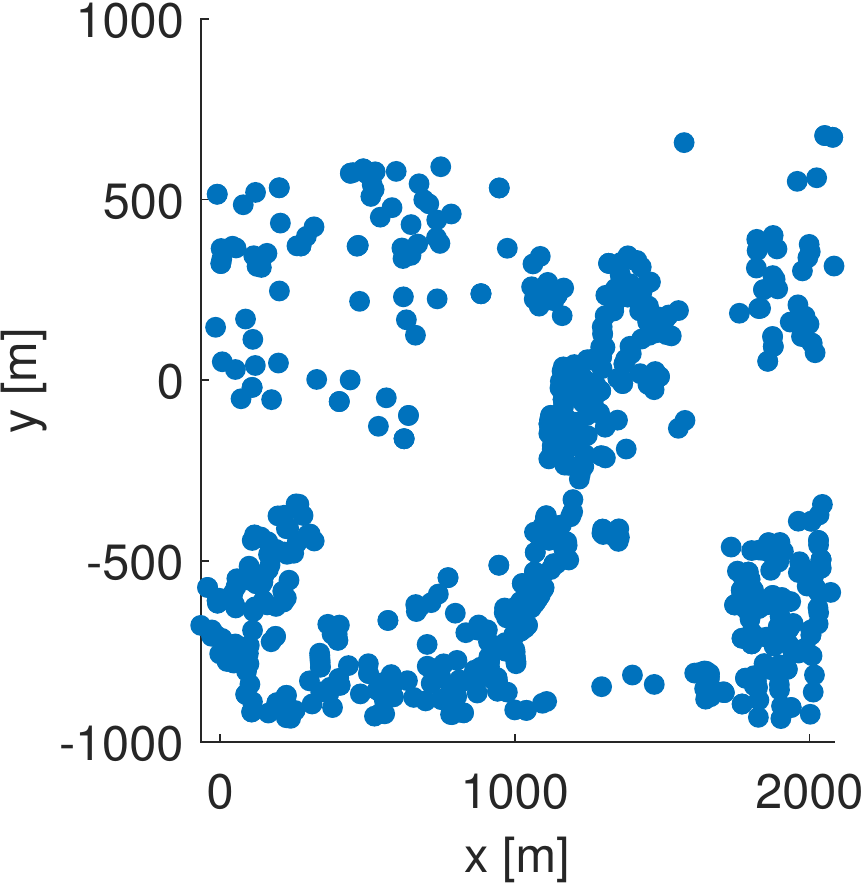}{0.22}{Delivery targets}
	\subfig{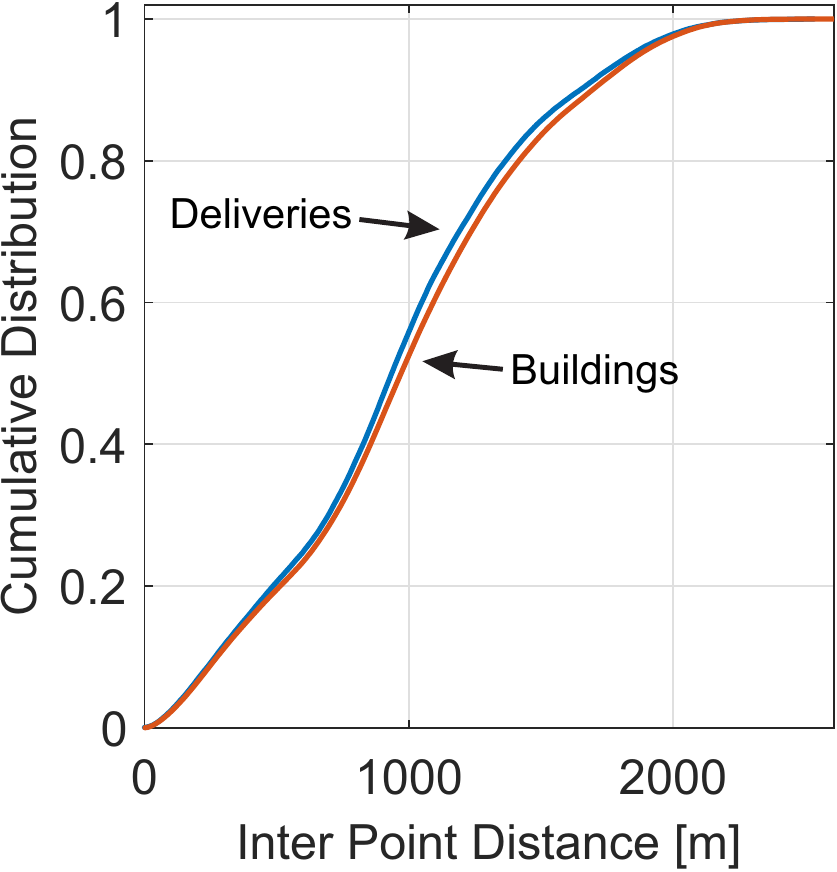}{0.22}{Delivery distribution}
	\caption{The spatial distribution of the deliveries over the scenario. The \ac*{IPD} of the deliveries being almost identical to the buildings' indicates an equal distribution of deliveries over the available buildings.}
	\label{fig:delivery_distribution}
\end{figure}

In this section, we present initial results for the proof-of-concept evaluation of different aspects of the overall delivery system.
%
%
First, the characteristics of the delivery sets which were generated for the evaluations are analyzed in Fig.\ref{fig:delivery_distribution}.
The delivery distribution nearly overlapping with the buildings' shows the deliveries were equally distributed over the available possible recipients in the scenario.

\subsection{Capacity and Time Requirements}

\begin{figure}[]
	\centering
	\subfig{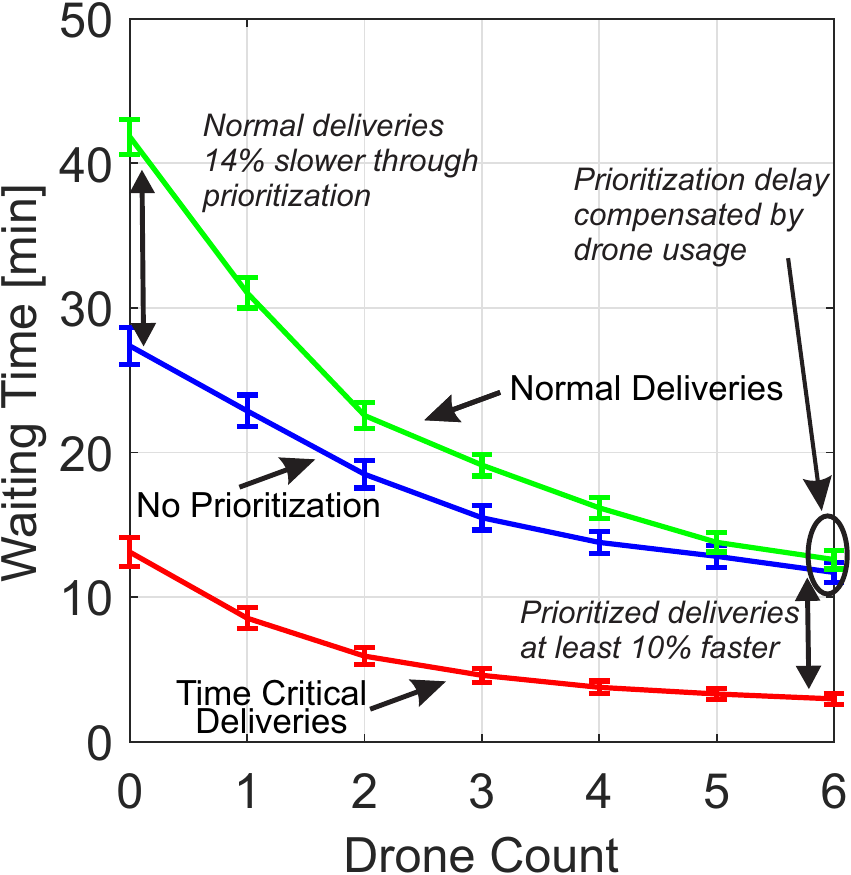}{0.22}{}
	\subfig{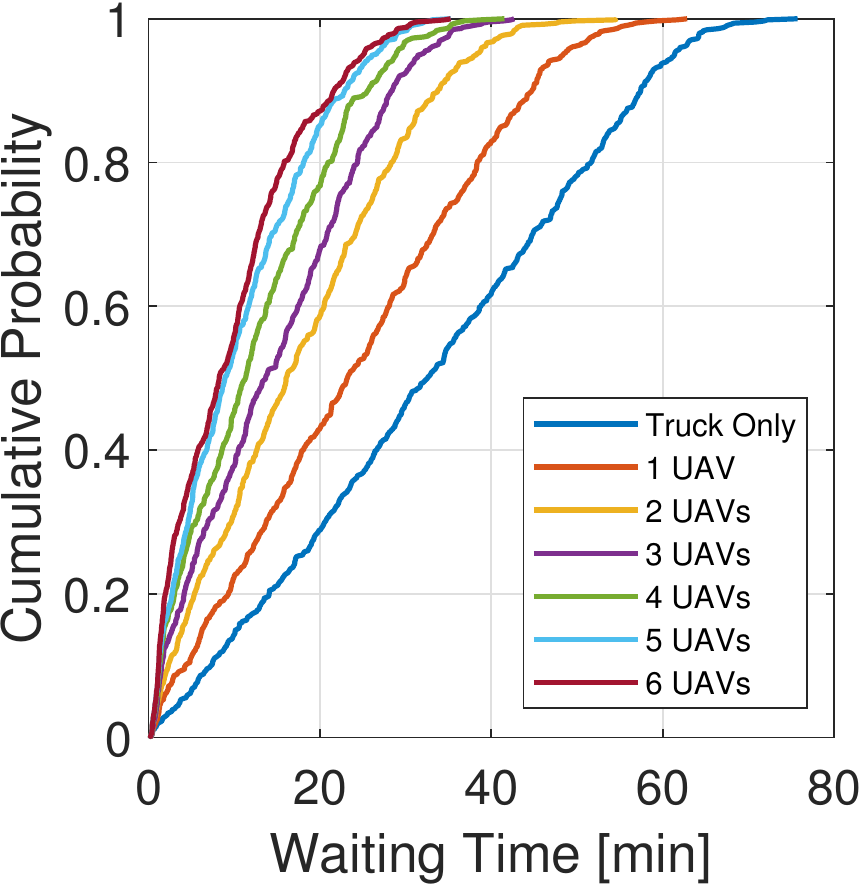}{0.22}{}

	\caption{Influence of the fleet drone count over waiting time of delivery categories and delivery capacity.}
	\label{fig:delivery-performance}
\end{figure}

The delivery performance with different fleet compositions ranging from only a truck to a hybrid fleet composed of a truck aided by up to 6 drones is shown in Fig.~\ref{fig:delivery-performance}.
%
%
The simple prioritization of medical deliveries shortens their waiting time by at least 50\%. This, however, delays normal deliveries. In our case study, the noticed waiting time increase for normal deliveries is up to 30\%. Furthermore, it shortens when an increasing number of drones is deployed to assist the truck and is almost compensated when six drones are used.
%
%
This variation in waiting times suggests an increase in delivery capacity, which we illustrated in Fig.~\ref{fig:delivery-performance} by analyzing the waiting time distribution with increased drone usage in the fleet. In our example, 80\% of the deliveries can be completed in under 20 minutes when the truck is assisted by 4 delivery drones.
%
%

\subsection{Suitable Network Approaches}

Next, we compare centralized and decentralized approaches in their suitability for connecting the delivery fleet and in their satisfaction of the requirements presented in Sec.~\ref{sec:communications}, for \ac{UAV} applications and especially drone aided parcel delivery. These are a maximal delay of 50 ms and nearly 99\% \ac{PDR}. \ac{LTE} is used as benchmark technology to represent \emph{centralized}  and coordinated medium access approaches. \ac{C-V2X} and \ac{WAVE}-based IEEE 802.11p are used as representatives of \emph{decentralized} medium access approaches.
The simulated traffic is a 100~ms periodic transmission of 190 Bytes \acp{CAM} to the delivery truck.

\begin{figure}[]
	\centering
	\subfig{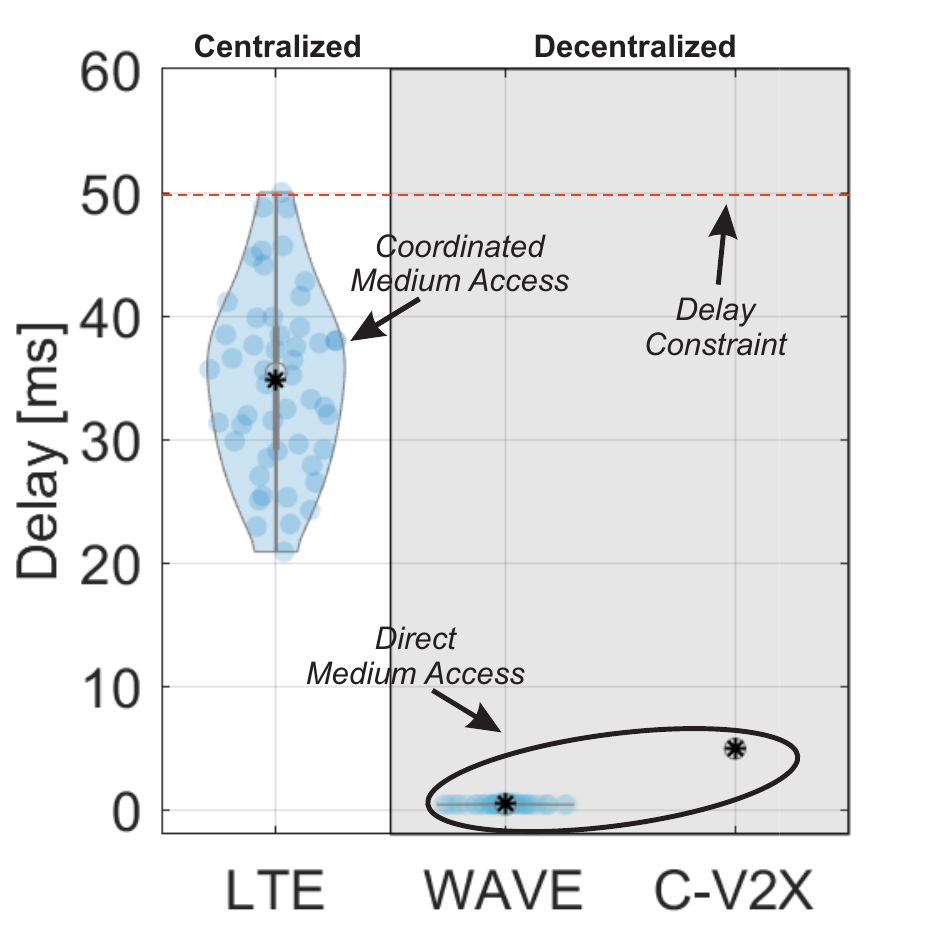}{0.22}{}
	\subfig{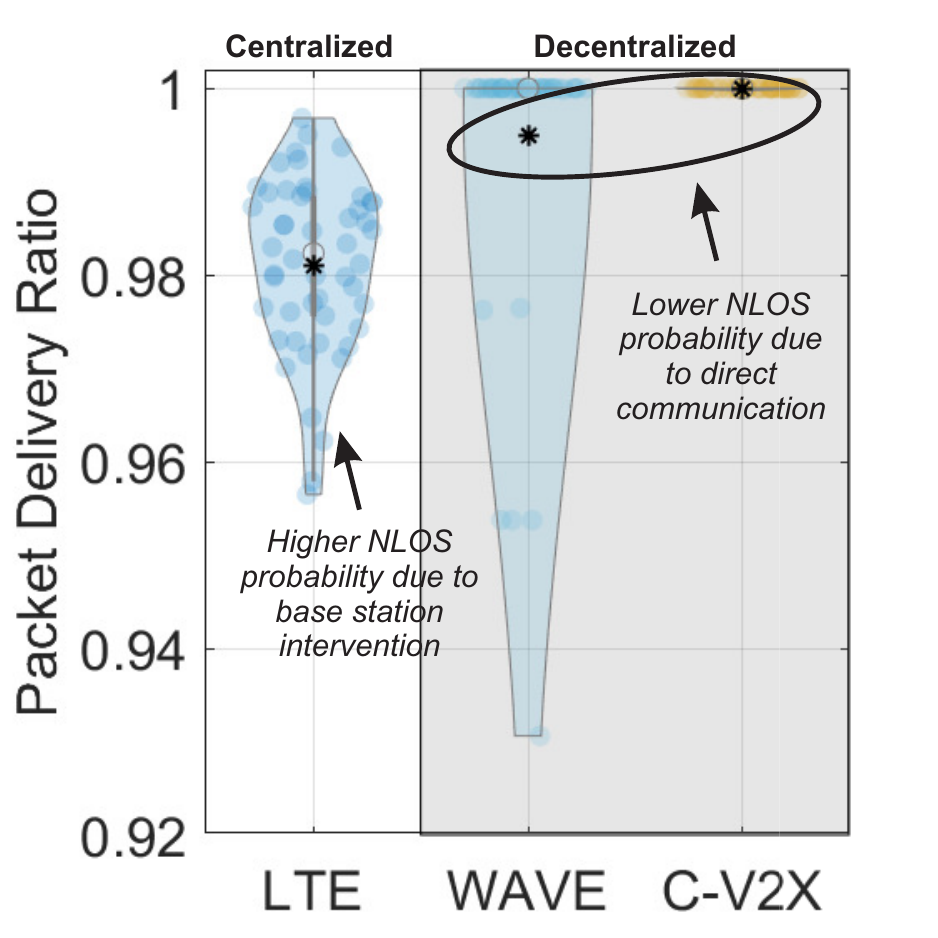}{0.22}{}

	\caption{Comparison of end-to-end metrics for different networking approaches for the delivery fleet.}
	\label{fig:networking}
\end{figure}

The results for latency and \ac{PDR} of the investigated communication technologies are shown in Fig.~\ref{fig:networking}. It can be seen that each technology can provide robust communication links. Still, a higher latency can be observed in the centralized approach represented by \ac{LTE}, which handles resource allocations centrally. The decentralized approaches, which implement direct medium access strategies, yield a smaller latency. The decentralized approaches also show more promising reliability with a \ac{PDR} very close to 1. The direct transmission path between sender and receiver gives a higher probability for \ac{LOS} situations than when using \ac{LTE}, where the base station - the \ac{eNB} - must be involved in the communication process. Newer \ac{LTE} releases may however perform better in this regard.
It can be further observed that \ac{C-V2X} offers slightly better reliability than \ac{WAVE}. This can be explained by the \ac{SPS}-based medium access which takes previous resource reservations into account to avoid resource conflicts in future reservation periods.

%
%

\section{Conclusion}

In this paper, we elaborate on the potential of \acp{UAV} to help meet the surging demand brought about by the sanitary measures taken to mitigate the COVID-19 pandemic. We presented a system concept to enhance existing last-mile delivery services with drone usage using advances in various research areas.
%
%
Within different case studies, we presented our groundwork results of various system aspects, namely the delivery capacity gains deriving from the enhanced mobility that hybrid delivery fleets of aerial and ground vehicles offer and the network communications requirements to operate such a fleet. A simple scheduling strategy is used to prioritize time-critical deliveries such as medical goods, and its effects on overall delivery performance are observed. Furthermore, a comparison between centralized and decentralized approaches for communications between hybrid delivery fleet members is carried out.
%
%
Similar to other services and applications (such as video conferencing), the unforeseen outbreak of the pandemic creates unexpected demands that will catalyze the development of solutions beyond the COVID-19 pandemic's scope.
In future work, we will focus on the development of scheduling algorithms to better balance short waiting times for time-critical deliveries with minimal additional delay to the normal ones while using a small number of drones. The development of communication-aware-mobility strategies is also considered.

\ifdoubleblind

\else

	\section*{Acknowledgment}
	
	\footnotesize
	Part of the work on this paper has been supported by the German Federal Ministry of Education and Research (BMBF) in the project A-DRZ (13N14857) as well as by the Deutsche Forschungsgemeinschaft (DFG) within the Collaborative Research Center SFB 876 ``Providing Information by Resource-Constrained Analysis'', projects A4 and B4.

\fi

\ifacm
\else
	\bibliographystyle{IEEEtran}
	\bibliography{Bibliography}
\fi

\end{document}